\documentstyle[12pt]{article}

\newcommand{\sect}[1]{\setcounter{equation}{0}\section{#1}}

\global\arraycolsep=1pt
\def\dirsl#1{\lefteqn{#1}/\,}
\def\dsl#1{{#1}\!\!\!\!/\,}

\begin{document}

\thispagestyle{empty}
\setcounter{page}{0}

\hfill MIT-CTP-2828

\hfill hep-th/9902129

\hfill February 18, 1999

\vspace{40pt}

\begin{center}
{\large {\bf KILLING SPINORS AND SUPERSYMMETRY ON AdS}}
\end{center}
\begin{center}
{\sl Eugene Shuster\footnote{\rm e-mail: eugeneus@mit.edu}}

{\it Center for Theoretical Physics, Massachusetts Institute of
Technology, Cambridge MA 02139, USA}
\end{center}

\vspace{40pt}


\begin{abstract}
In this paper we consturct several supersymmetric theories on AdS$_{5}$
background. We discuss the proper definition of the Killing equation
for the symplectic Majorana spinors required in AdS$_{5}$ supersymmetric
theories. We find that the symplectic Killing spinor
equation involves a matrix $M$ in the USp($2N$) indices whose role was
not recognized previously. Using the correct Killing spinors we
explicitly confirm that the particle masses in the constructed theories
agree with the predictions of the AdS/CFT correspondence. Finally, we
establish correct O($d-1,2$) isometry transformations required to keep
the Lagrangian invariant on AdS$_{d}$.
\end{abstract}
\vfill\eject

\sect{Introduction}

Recent work on the AdS/CFT correspondence \cite{maldacena, klebanov,
witten} has brought renewed interest
in the subject of supersymmetric field theory in anti-de Sitter space,
particularly for AdS$_{5}$. We have found that several basic questions
are not clearly discussed in the literature, and it is our aim to clarify
them in the present paper. These questions include the proper Killing
equation for the symplectic Majorana spinor required in AdS$_{5}$
SUSY, and the Lagrangian and transformation rules for the SU($N$) gauge
multiplet, the conformal scalar multiplet, and the massive scalar
multiplet.

It was a surprise to us that the symplectic Killing spinor equation
involves a matrix $M$ in the USp($2N$) indices whose role was not
recognized previously. It turns out that $M$ also enters the
transformation rules and the Lagrangian of the basic supermultiplets. In
this paper we will develop a full description of these basic
supermultiplets on AdS$_{5}$ using the properly defined symplectic
Killing spinors.

In the body of the paper, we will work with a metric of $(+,-, \cdots
,-)$ signature unless stated otherwise:
\begin{equation} \label{adsmet}
ds^{2} = e^{2 a r} \eta_{\alpha \beta} dx^{\alpha} dx^{\beta} - dr^{2}.
\end{equation}
With this choice of metric, the Ricci curvature is $R_{\mu\nu} = (d - 1) a^{2}
g_{\mu\nu}$. We give a summary of results for the $(-,+, \cdots ,+)$
signature in Appendix A. It is our hope that the results in
this paper will prove useful for developing further understanding of
physics on AdS$_{5}$.

\sect{Killing spinors on AdS}

It is known that in $d = 5, 6, 7$ mod 8 regular Majorana fermions
cannot be defined \cite{pvan}. Instead, in these dimensions we can
define symplectic Majorana fermions which are spinors satisfying the
following condition \cite{pvan,cremmer,gunaydin}:
\begin{equation} \label{symmaj}
\chi^{i} = C (\overline{\chi}^{i})^{T}
\end{equation}
where $C$ is the charge conjugation matrix and 
\begin{equation} \label{bar}
\overline{\chi}^{i} \equiv \chi_{i}^{\dagger} \gamma_{0}.
\end{equation}
In general, $i = 1, 2, \ldots ,
2n$ and the indices are raised and lowered with a symplectic metric
$\Omega_{ij}$ which obeys
\begin{equation}
\Omega^{T} = - \Omega, ~ \Omega \Omega^{\star} = - I.
\end{equation}
In this paper, we will only be interested in a pair of symplectic
Majorana spinors, so $i = 1,2$ and 
\begin{equation}
\epsilon = \left( \begin{array}{cc} 0 & 1 \\ -1 & 0 \end{array} \right)
\end{equation}
is the symplectic metric which will be used throughout this paper. To
simplify our computations, we will only use objects with all the
symplectic indices lowered by inserting the symplectic metric
explicitly, for example, $\chi^{i} = \chi_{j} \epsilon^{j i} = -
\epsilon_{i j} \chi_{j}$. Note that in our convention,
\begin{equation}
\epsilon^{i j} = - \epsilon^{j i} = \epsilon_{i j}.
\end{equation}
Because of definition (\ref{bar}), we will treat $\overline{\chi}^{i}$
as an object with its symplectic index down, but will at times
employ $\overline{\chi}^{i}$ notation to save space.

For the rest of this paper we will work in $d = 5$ unless explicitly
stated otherwise. Because now the Clifford algebra contains
$\gamma_{5}$, the Fierz transformations become simpler:
\begin{equation}
\left(\begin{array}{l} s \\ v \\ t \end{array} \right)(4,2;3,1) = - \frac{1}{4}
\left[ \begin{array}{ccc} 1 & 1 & 1 \\ 5 & -3 & 1 \\ 10 & 2 & -2
\end{array} \right] \left(\begin{array}{l} s \\ v \\ t \end{array}
\right)(4,1;3,2)
\end{equation}
where
\begin{equation}
\begin{array}{l}
s(a,b;c,d) = \overline{\psi}_{a} \psi_{b} \overline{\psi}_{c} \psi_{d} \\
v(a,b;c,d) = \overline{\psi}_{a} \gamma_{\mu} \psi_{b}
\overline{\psi}_{c} \gamma^{\mu} \psi_{d} \\
t(a,b;c,d) = -\frac{1}{2} \overline{\psi}_{a} \gamma_{\mu\nu} \psi_{b}
\overline{\psi}_{c} \gamma^{\mu\nu} \psi_{d}.
\end{array}
\end{equation}
Another useful identity to keep in mind in 5 dimensions is
the symplectic Majorana flip formula \cite{cremmer}
\begin{equation}
\overline{\chi}^{i} \gamma_{\mu_{1}} \gamma_{\mu_{2}} \ldots
\gamma_{\mu_{n-1}} \gamma_{\mu_{n}} \psi^{j} = \overline{\psi}^{j}
\gamma_{\mu_{n}} \gamma_{\mu_{n-1}} \ldots \gamma_{\mu_{2}}
\gamma_{\mu_{1}} \chi^{i} 
\end{equation}
which written in our notation becomes
\begin{equation}
\chi_{i}^{\dagger} \gamma_{0} \gamma_{\mu_{1}} \gamma_{\mu_{2}} \ldots
\gamma_{\mu_{n-1}} \gamma_{\mu_{n}} \psi_{j} = -\epsilon_{i l}
\epsilon_{j k} \psi_{k}^{\dagger} \gamma_{0} \gamma_{\mu_{n}}
\gamma_{\mu_{n-1}} \ldots \gamma_{\mu_{2}} \gamma_{\mu_{1}} \chi_{l}.
\end{equation}
This formula comes about because the charge conjugation matrix, $C$, is
such that 
\begin{equation} \label{chargeconj}
C \gamma_{\mu} C^{-1} = \gamma_{\mu}^{T},
\end{equation}
which is different from 4 dimensions where there is a minus sign on the
right hand side of the equation. Because of that minus sign, a Majorana
Killing spinor equation in 4 dimensions can be defined in a
straightforward manner 
\cite{burges}:
\begin{equation} \label{kilspdirac}
D_{\mu} \epsilon = i \frac{a}{2} \gamma_{\mu} \epsilon.
\end{equation}
Note that because this equation satisfies the Ricci identity (see
eq.~(\ref{ricci}) below), it can be interpreted as a Killing equation in
arbitrary dimension for a complex unconstrained spinor, which was studied
previously \cite{lupope}. An extension of the above definition to 5
dimensions fails because the left hand side satisfies the symplectic
Majorana condition (\ref{symmaj}) while the right hand side of the above
equation does not, due to eq.~(\ref{chargeconj}). This led us to
consider a generalized form of the Killing equation for the symplectic
Majorana Killing spinors:
\begin{equation} \label{kilspeq}
D_{\mu} \epsilon_{i} = i M_{i j} \frac{a}{2} \gamma_{\mu} \epsilon_{j},
\end{equation}
where $M_{i j}$ is an unknown $2 \times 2$ matrix, and
$D_{\mu} \psi = (\partial_{\mu} + \frac{1}{2} \omega_{\mu a b} \sigma^{a
b}) \psi$. It is important to note that this form of the symplectic
Killing spinor equation and the subsequent supersymmetry
transformations stemming from it are compatible with AdS$_{5}$
supergravity transformation rules \cite{gunaydin}, although Killing
spinors were not discussed there. We obtain the properties of the matrix
$M_{i j}$ by applying the Ricci identity and the symplectic Majorana
condition to eq.~(\ref{kilspeq}). Using eqs.~(\ref{symmaj}) and
(\ref{chargeconj}) yields a condition on  $M_{ij}$, which written in the
matrix form becomes 
\begin{equation} \label{M1}
M = \epsilon M^{\star} \epsilon,
\end{equation}
where $\epsilon$ is the symplectic metric. On the other hand, Ricci
identity yields 
\begin{equation} \label{ricci}
[D_{\mu}, D_{\nu}] \epsilon_{i} \equiv \frac{1}{2} R_{\mu \nu a b} \sigma^{a
b} \epsilon_{i} = a^{2} \sigma_{\mu \nu} \epsilon_{i} = a^{2} \sigma_{\mu
\nu} (M^{2})_{i j} \epsilon_{j},
\end{equation}
that is
\begin{equation} \label{M2}
M^{2} = 1.
\end{equation}
Putting equations~(\ref{M1}) and (\ref{M2}) together, we can easily
obtain the most general form of the matrix $M$:
\begin{equation} \label{M}
M = \left( \begin{array}{ll} \cos\theta & \sin\theta e^{-i \phi} \\
\sin\theta e^{i \phi} & -\cos\theta \end{array} \right) = \vec{x}
\cdot \vec{\sigma}
\end{equation}
where $\theta$ and $\phi$ are angles taking values between 0 and $2 \pi$,
and \[ \vec{x} = (\sin\theta \cos\phi, \sin\theta \sin\phi, \cos\theta), \]
and $\vec{\sigma} = (\sigma_{1}, \sigma_{2}, \sigma_{3})$ is a vector of
Pauli matrices. Hence, matrix $M$ can be interpreted as an
element of the Lie algebra of USp(2). Furthermore, it is easily seen that
the Killing spinor equation~(\ref{kilspeq}) is USp(2)-covariant: take
$M$ to be an allowed matrix appearing in eq.~(\ref{kilspeq}),
then USp(2) rotate the Killing spinors, $\epsilon_{i}^{\prime} = U_{i j}
\epsilon_{j}$, and write new Killing spinor equation for the rotated
spinors -- we obtain the same form of Killing spinor equation but with a
different matrix $M^{\prime} = U^{\dagger} M U$ which satisfies both
conditions~(\ref{M1}) and (\ref{M2}), hence giving us a valid Killing
spinor equation.

Complex Killing spinors on $d$-dimensional anti-de Sitter spacetime
AdS$_{d}$ for arbitrary $d$ have been constructed before
\cite{lupope,lupoper}. They are solutions of eq.~(\ref{kilspdirac}) and
take the form (see Appendix A for conversion between signatures)
\begin{equation} \label{kilspn}
\epsilon = e^{\frac{i}{2} a r \gamma_{r}} \left( 1 + \frac{i}{2}
a x^{\alpha} \gamma_{\alpha} (1 - i \gamma_{r}) \right) \epsilon_{0}.
\end{equation}
Solution of equation~(\ref{kilspeq}) can easily be obtained from
the solution~(\ref{kilspn}) by substituting $M_{ij} \gamma_{\mu}$ for
every $\gamma_{\mu}$ in eq.~(\ref{kilspn}):
\begin{equation} \label{kilspM}
\epsilon_{i} = \left(e^{\frac{i}{2} a r M \gamma_{r}}\right)_{ij} \left(
\delta_{jk} + \frac{i}{2} a x^{\alpha} \gamma_{\alpha} (M_{jk} - i
\delta_{jk} \gamma_{r}) \right) \xi_{k} 
\end{equation}
where $\xi_{j}$ is a pair of constant symplectic Majorana spinors.

Now, for each matrix $M$ above, let us construct a Dirac spinor as a
linear combination of the two symplectic Majorana spinors . Assume the
most general relation between two symplectic Majorana spinors and a
Dirac spinor:
\begin{equation} \label{dirac}
\psi = A \epsilon_{1} + B \epsilon_{2}
\end{equation}
with the unknown coefficients $A$ and $B$. To be consistent,
eq.~(\ref{dirac}) should produce the right equation for the Dirac
Killing spinors, eq.~(\ref{kilspdirac}), when combined with equation for
the symplectic Majorana Killing spinors, eq.~(\ref{kilspeq}). Using
eq.~(\ref{M}) we find that this condition is satisfied by the following
normalized Dirac spinor 
\begin{equation} \label{norm-dirac}
\psi =  e^{i \frac{\phi}{2}} \cos \frac{\theta}{2} ~ \epsilon_{1} + e^{-i
\frac{\phi}{2}} \sin \frac{\theta}{2} ~ \epsilon_{2},
\end{equation}
where we could also choose $-$ instead of $+$ between the two
terms. This expression will be useful later in the paper.

Finally, it's worth noting a general form of the matrix $M$ for more
than 2 spinors. In the case of $2n$ spinors, $M$ is a $2n \times 2n$
matrix which takes a block form
\begin{equation}
M = \left( \begin{array}{cc} A & B \\ B^{\star} & -A^{\star}
\end{array}\right)
\end{equation}
where $A$ and $B$ are $n \times n$ complex matrices which satisfy the
following equations:
\begin{equation}
\begin{array}{l}
AB = B A^{\star} \\ A^{2} + B B^{\star} = I.
\end{array}
\end{equation}
It is easy to see that for $n=1$, above equations yield precisely the
matrix $M$ given in equation~(\ref{M}).

\sect{The on-shell USp(2) supersymmetric U(1) Yang-Mills theory on
AdS$_{5}$}

Let us start with the massless USp(2) supersymmetric Yang-Mills theory in flat
4+1 spacetime. SU(2) version of this theory has been developed by Zizzi
\cite{zizzi}. U(1) theory is easily obtained from SU(2) theory:
\begin{equation} \label{flatYM}
{\cal L} = - \frac{1}{4} F^{\mu\nu} F_{\mu\nu} + \frac{1}{2} D_{\mu}\phi
D^{\mu}\phi + \frac{i}{2} \overline{\chi}^{i} \dsl{D} \chi_{i}
\end{equation}
and invariant under the following supersymmetry transformations:
\begin{equation} \begin{array}{l}
\delta A_{\mu} = i \overline{\eta}^{i} \gamma_{\mu} \chi_{i} \\
\delta \phi = i \overline{\eta}^{i} \chi_{i} \\
\delta \chi_{i} = (\sigma_{\mu\nu} F^{\mu\nu} - \dsl{D} \phi) \eta_{i}
\end{array} \end{equation}
where $\mu,\nu = 0, \ldots ,4$ and $i=1,2$. To describe the same theory on
AdS$_{5}$ not only do we need to have additional terms in the
supersymmetry transformations but we will have nonzero mass terms for
both the scalar $\phi$ and the spinors $\chi_{i}$ for the case of
massless gauge potential. In fact, compactification of ${\cal N} = 2$
supergravity on S$^{5}$ \cite{kimromans} or AdS/CFT correspondence
\cite{witten,sweedes} let us determine these masses:
\begin{equation} \label{mass}
m^{2}(A_{\mu}) = 0, ~ \mid m(\psi)\mid = \frac{1}{2}, ~ m^{2}(\phi) = -4.
\end{equation}
Hence, the U(1) Yang-Mills theory on AdS$_{5}$ should be the flat U(1)
theory~(\ref{flatYM}) plus the above mass terms:
\begin{equation} \label{adsYM}
{\cal L} = - \frac{1}{4} F^{\mu\nu} F_{\mu\nu} + \frac{1}{2} D_{\mu}\phi
D^{\mu}\phi + \frac{i}{2} \overline{\chi}^{i} \dsl{D} \chi_{i} - a \mu_{ij}
\overline{\chi}^{i} \chi_{j} - \frac{1}{2} a^{2} m^{2} \phi^{2}.
\end{equation}
Using the proper Killing spinor equation~(\ref{kilspeq}) for the
symplectic Majorana spinors, we find that theory~(\ref{adsYM}) is
invariant under the supersymmetry transformations
\begin{equation} \label{susygauge}
\begin{array}{l}
\delta A_{\mu} = i \overline{\eta}^{i} \gamma_{\mu} \chi_{i} \\
\delta \phi = i \overline{\eta}^{i} \chi_{i} \\
\delta \chi_{i} = (\sigma_{\mu\nu} F^{\mu\nu} - \dsl{D} \phi) \eta_{i} -
2 i a \phi M_{ij} \eta_{j} 
\end{array}
\end{equation}
where $M_{ij}$ is the matrix given by eq.~(\ref{M}). Furthermore,
supersymmetry determines the values of the masses in eq.~(\ref{adsYM}):
\begin{equation} 
\mu = - \frac{1}{4} M, ~ m^{2} = - 4.
\end{equation}

It is easy to show that given a definition of a properly
normalized Dirac spinor as in eq.~(\ref{norm-dirac}),
\begin{equation} \label{kin-dirac}
\frac{i}{2} \overline{\chi}^{i} \dsl{D} \chi_{i} = i
\overline{\psi} \dsl{D} \psi
\end{equation}
and
\begin{equation} \label{mass-dirac}
\frac{1}{2} M_{ij} \overline{\chi}^{i} \chi_{j} = \overline{\psi} \psi.
\end{equation}
Hence, this theory contains a Dirac spinor of mass equal to
$\frac{1}{2}$ and one real scalar of mass equal to $-4$. These
masses agree completely with our previous predictions given in
equation~(\ref{mass}). 

To complete the description of this theory we need to write down the
supersymmetry algebra. Using eq.~(\ref{susygauge}), we find that
\begin{equation}
[\delta_{1},\delta_{2}] \phi = 2 i \overline{\eta}^{i}_{1} \gamma^{\mu}
\eta_{2 i} D_{\mu} \phi
\end{equation}
and similarly a usual expression for $[\delta_{1},\delta_{2}] A_{\mu}$
(up to equations of motion and gauge transformations) because just like
in the scalar case above all the terms proportional to $a$
cancel. However, supersymmetry algebra for the spinors is more
interesting:
\begin{equation} \label{susyYM}
[\delta_{1},\delta_{2}] \chi_{i} = 2 i D_{\mu} \chi_{i}
\overline{\eta}^{j}_{1} \gamma^{\mu} \eta_{2 j} + 3 a M_{ij} \chi_{j}
\overline{\eta}^{k}_{1} \eta_{2 k} + \frac{a}{2} \gamma^{\mu\nu}
\chi_{i} M_{kj} \overline{\eta}^{k}_{1} \gamma_{\mu\nu} \eta_{2 j},
\end{equation}
where we used spinor equations of motion
\begin{equation}
\dsl{D} \chi_{i} = \frac{i}{2} a M_{ij} \chi_{j}
\end{equation}
and the following useful identities
\begin{equation} \begin{array}{l}
M_{nl} \delta_{ij} - M_{il} \delta_{nj} + M_{ij} \delta_{nl}- M_{nj}
\delta_{il} = 0 \\
\epsilon_{nj} (M \epsilon)_{ik} + \delta_{kn} M_{ij} = M_{in}
\delta_{jk} \\
\epsilon_{nj} \epsilon_{ik} + \delta_{kn} \delta_{ij} = \delta_{in}
\delta_{jk}.
\end{array} \end{equation}
To explain the terms appearing in this algebra, first consider only the
fermionic part of the Lagrangian 
\begin{equation}
{\cal L}_{F} = \frac{i}{2} \overline{\chi}^{i} \dsl{D}
\chi_{i} + \frac{1}{4} a M_{ij} \overline{\chi}^{i} \chi_{j}.
\end{equation}
From the properties of the matrix $M$ (eqs.~(\ref{M1}) and (\ref{M2})), it
follows that there is an additional U(1) symmetry in the theory:
\begin{equation} \label{extraspinsym}
\delta \chi_{i} = i M_{ij} \chi_{j}.
\end{equation}
This extra symmetry manifests itself in the supersymmetry algebra, as we
see from the second term in eq.~(\ref{susyYM}). Furthermore,
supersymmetry algebra involves a term proportional to 
\begin{equation} \label{extraalgterm}
\gamma^{\mu\nu} \chi_{i} M_{kj} \overline{\eta}^{k}_{1} \gamma_{\mu\nu}
\eta_{2 j} 
\end{equation}
which at first glance appears unusual. However, there is a clear and
dimension independent explanation of this term. The proof of the
following arguments is presented in Appendix B. Below, we chose to
work in 4 dimensions because we do not wish to involve the symplectic
indices. In 5 dimensions, the following discussion is slightly more
involved but follows the same outline as the proof in Appendix B. To
facilitate the explanation, let us look at the AdS$_{4}$ supersymmetric theory
\cite{burges}. If we compute supersymmetry algebra of the fermions using
transformations (3.1) of that paper, we obtain (with our conventions)
\begin{equation} \label{sigmasigma}
[\delta_{1},\delta_{2}] L \psi = i D^{\mu} L \psi
\overline{\epsilon}_{1} \gamma_{\mu} \epsilon_{2} + \frac{a}{4}
\gamma^{\mu \nu} L \psi \overline{\epsilon}_{1} \gamma_{\mu \nu} \epsilon_{2}.
\end{equation}
This algebra contains an ``extra'' term of the same form as
eq.~(\ref{extraalgterm}). The explanation of this ``extra'' term in 
the algebra lies in the fact that the naive isometry transformation
\begin{equation} 
\delta \psi = K^{\mu} D_{\mu} \psi,
\end{equation}
where $K^{\mu} = i \overline{\epsilon}_{1} \gamma^{\mu} \epsilon_{2}$ is
an O(3,2) Killing vector, is actually not a symmetry of the kinetic term
in curved space. On a curved manifold, we need to add more terms to this
variation because $D_{\mu} K_{\nu}$ no longer equals to 0. In
particular, in AdS we need to add precisely the ``extra term'' in
eq.~(\ref{sigmasigma}) in order to recover a symmetry of the
Lagrangian. Using the fact that 
\begin{equation}
D_{\mu} K_{\nu} = a \overline{\epsilon}_{1} \gamma_{\mu \nu} \epsilon_{2}
\end{equation}
we expect that in AdS$_{4}$ the full O(3,2) isometry requires the
following transformation rule:
\begin{equation} \label{adstransl}
\delta \psi = i D^{\mu} \psi \overline{\epsilon}_{1} \gamma_{\mu}
\epsilon_{2} + a \sigma^{\mu \nu} \psi \overline{\epsilon}_{1}
\sigma_{\mu \nu} \epsilon_{2} = K^{\mu} D_{\mu} \psi + \frac{1}{4}
D^{\mu} K^{\nu} \gamma_{\mu \nu} \psi
\end{equation}
which can be verified to be a symmetry of the Lagrangian (see Appendix
B). Hence, we indeed expect a term like (\ref{extraalgterm}) in the
supersymmetry algebra of our theory on AdS$_{5}$.

Let us finally note that extending the above results to an SU($N$)
gauge theory is quite trivial. Assuming that all the matter fields are
in the adjoint representation of the gauge group SU($N$), the Lagrangian
is
\begin{equation} \label{adsYMN}
{\cal L} = - \frac{1}{4} F^{\mu\nu a} F_{\mu\nu}^{a} + \frac{1}{2}
D_{\mu}\phi^{a} D^{\mu}\phi^{a} + \frac{i}{2} \overline{\chi}^{i a}
\dsl{D} \chi_{i}^{a} + \frac{1}{4} a M_{ij} \overline{\chi}^{i a}
\chi_{j}^{a} + 2 a^{2} \phi^{a} \phi^{a} - \frac{i}{2} g f_{abc}
\overline{\chi}^{i a} \phi^{b} \chi_{i}^{c}
\end{equation}
where $a, b, c = 1, 2, \ldots , N^{2}-1$ and the covariant derivatives
are now defined as
\begin{equation} \begin{array}{l}
D_{\mu} \psi_{i}^{a} = \partial_{\mu} \psi_{i}^{a} + \frac{1}{2}
\omega_{\mu\nu\rho} \sigma^{\nu\rho} \psi_{i}^{a} + g f_{abc}
A_{\mu}^{b} \psi_{i}^{c} \\
D_{\mu} \phi^{a} = \partial_{\mu} \phi^{a} + g f_{abc} A_{\mu}^{b}
\phi^{c} \\
F_{\mu\nu}^{a} = \partial_{\mu} A_{\nu}^{a} - \partial_{\nu} A_{\mu}^{a}
+ g f_{abc} A_{\mu}^{b} A_{\nu}^{c}.
\end{array} \end{equation}
With these definitions, the theory (\ref{adsYMN}) is invariant under
exactly the same transformations as before, keeping in mind the
definitions above:
\begin{equation}
\begin{array}{l}
\delta A_{\mu}^{a} = i \overline{\eta}^{i} \gamma_{\mu} \chi_{i}^{a} \\
\delta \phi^{a} = i \overline{\eta}^{i} \chi_{i}^{a} \\
\delta \chi_{i}^{a} = (\sigma_{\mu\nu} F^{\mu\nu a} - \dsl{D} \phi^{a})
\eta_{i} - 2 i a \phi^{a} M_{ij} \eta_{j}.
\end{array}
\end{equation}
Therefore, all the results, including the field masses and the
supersymmetry algebra, remain exactly the same in the case of SU($N$)
gauge theory.

\sect{The on-shell USp(2) supersymmetric conformal scalar theory on
AdS$_{5}$}

This theory describes 2 massless symplectic Majorana fermions and 4
massive real scalar fields with the same mass for all 4 scalars. A
generalization of this theory has been developed in flat 5-dimensional
spacetime \cite{sharpe}. As in the previous section, we use the flat
theory to develop this same theory on AdS$_{5}$:
\begin{equation} \label{adsconf}
{\cal L} = \frac{1}{2} \partial_{\mu} \phi^{I} \partial^{\mu} \phi^{I} +
\frac{i}{2} \overline{\lambda}^{i} \dsl{D} \lambda_{i}
- \frac{1}{2} a^{2} m^{2} \phi^{I} \phi^{I}
\end{equation}
where $\mu = 0, \ldots ,4$, $I = 1, \ldots ,4$, and $i = 1,2$. The
theory~(\ref{adsconf}) is invariant under the supersymmetry
\begin{equation} \label{susyconf}
\begin{array}{l}
\delta \phi^{I} = (\sigma^{I} \epsilon)_{ij} \overline{\epsilon}^{i}
\lambda_{j} \\
\delta \lambda_{i} = i (\epsilon \sigma^{I})_{ji} \dirsl{\partial}
\phi^{I} \epsilon_{j} + a \frac{3}{2} ({\sigma^{I}}^{T} \epsilon M)_{ij}
\epsilon_{j} \phi^{I}
\end{array}
\end{equation}
where $\sigma^{I}=(\vec{\sigma}, i {\bf 1})$, $M$ is the matrix found in
eq.~(\ref{M}), $\epsilon$ is the symplectic metric in 
the matrix form, and $\epsilon_{i}$ is a symplectic Majorana Killing
spinor. Supersymmetry also determines the value of $m$ in this theory:
\begin{equation}
m^{2} = - \frac{15}{4}.
\end{equation}
Note that this formula agrees with the mass formula for the conformally
coupled scalar field in 5 dimensions \cite{burges,mezincescu}.

Now, we are ready to compute the supersymmetry algebra for this
theory. After a short computation, we obtain
\begin{equation} \label{susyalgsc}
[\delta_{1},\delta_{2}] \phi^{I} = 2 i \overline{\epsilon}_{1}^{i}
\gamma^{\mu} \epsilon_{2 i} \partial_{\mu} \phi^{I} + a \frac{3}{2}
\overline{\epsilon}_{2}^{i} \left(\sigma^{I} {\sigma^{J}}^{\dagger} M - M
\sigma^{J} {\sigma^{I}}^{\dagger}\right)_{ij} \epsilon_{1 j} \phi^{J}.
\end{equation}
It appears (and can be confirmed by an explicit computation) that there
exists an ``extra'' symmetry in this theory
\begin{equation} \label{extraO4}
\delta \phi^{I} = \overline{\epsilon}_{2}^{i} (\sigma^{I}
{\sigma^{J}}^{\dagger} M - M \sigma^{J} {\sigma^{I}}^{\dagger})_{ij}
\epsilon_{1 j} \phi^{J}.
\end{equation}
In fact, this transformation represents rotation of the scalar fields
$\delta \phi^{I} = i \overline{\epsilon}_{2}^{i} \epsilon_{1 i} T^{IJ}
\phi^{J}$ where $T$ is a $4 \times 4$ matrix
\begin{equation} \label{T}
T = \left( \begin{array}{cccc} 0 & x_{3} & -x_{2} & -x_{1} \\ -x_{3} & 0 &
x_{1} & -x_{2} \\ x_{2} & -x_{1} & 0 & -x_{3} \\ x_{1} & x_{2} & x_{3} & 0
\end{array} \right)
\end{equation}
and $\vec{x}$ is defined in eq.~(\ref{M}). Hence, for each fixed matrix
$M$, this is a particular representation of the SO(2) subgroup of the
obvious SO(4) symmetry of the scalar Lagrangian.

The spinor algebra, on the other hand, presents nothing new, although
the previous extra term associated with the O(4,2) isometry does. In
order to calculate this algebra, we need to know a few useful identities
given below:
\begin{equation} \label{Msigmaid}
\begin{array}{l}
(\epsilon \sigma^{I})_{ji} (\sigma^{I} \epsilon)_{mn} - \sigma^{I}_{mi}
{\sigma^{I}_{jn}}^{\star} = - 4 \delta_{jm} \delta_{in} \\
(\epsilon \sigma^{I})_{ji} (\sigma^{I} \epsilon)_{mn} + \sigma^{I}_{mi}
{\sigma^{I}_{jn}}^{\star} = 0 \\
(\epsilon \sigma^{I})_{ji} (M \sigma^{I} \epsilon)_{mn} + \sigma^{I}_{mi}
(M \sigma^{I})_{jn}^{\star} = 0 \\
(\epsilon \sigma^{I})_{ji} (M \sigma^{I} \epsilon)_{mn} - \sigma^{I}_{mi}
(M \sigma^{I})_{jn}^{\star} = - 4 \delta_{in} M_{mj} \\
(\epsilon M \sigma^{I})_{ji} (\sigma^{I} \epsilon)_{mn} - (M \sigma^{I})_{mi}
{\sigma^{I}_{jn}}^{\star} = 0 \\
(\epsilon M \sigma^{I})_{ji} (\sigma^{I} \epsilon)_{mn} + (M \sigma^{I})_{mi}
{\sigma^{I}_{jn}}^{\star} = 4 M_{mj} \delta_{in} \\
(M \epsilon {\sigma^{I}}^{\dagger})_{ij} (\sigma^{I} \epsilon)_{mn} + (M
{\sigma^{I}}^{T})_{im} {\sigma^{I}_{jn}}^{\star} = 0 \\
(M \epsilon {\sigma^{I}}^{\dagger})_{ij} (\sigma^{I} \epsilon)_{mn} - (M
{\sigma^{I}}^{T})_{im} {\sigma^{I}_{jn}}^{\star} = - 4 \delta_{jm} M_{in}.
\end{array}
\end{equation}
Then, using these identities we arrive at the following result
\begin{equation} \label{susyalgsp}
[\delta_{1},\delta_{2}] \lambda_{i} = 2 i D_{\mu} \lambda_{i}
\overline{\epsilon}_{1}^{j} \gamma^{\mu} \epsilon_{2 j} + \frac{a}{2}
\gamma^{\mu \nu} \lambda_{i} M_{kj} \overline{\epsilon}_{1}^{k} 
\gamma_{\mu \nu} \epsilon_{2 j}
\end{equation}
which is remarkably similar to the supersymmetric algebra for the spinors
in AdS$_{4}$ as given by eq.~(\ref{sigmasigma}).

\sect{The on-shell USp(2) supersymmetric massive scalar theory on
AdS$_{5}$}

Similarly to the conformal scalar theory presented in the previous
section, this theory describes 2 massive symplectic Majorana fermions
and 4 massive real scalar fields. The theory is described by an action
similar to that of conformal scalar theory given by eq.~(\ref{adsconf}):
\begin{equation} \label{adsmass}
{\cal L} = \frac{1}{2} \partial_{\mu} \phi^{I} \partial^{\mu} \phi^{I} +
\frac{i}{2} \overline{\lambda}^{i} \dsl{D} \lambda_{i} - \frac{1}{2} a
\mu M_{ij} \overline{\lambda}^{i} \lambda_{j} - \frac{1}{2} a^{2}
m^{2}_{IJ} \phi^{I} \phi^{J}.
\end{equation}
The new feature here is a symmetric real $4 \times 4$ matrix
$m^{2}_{IJ}$ which is not assumed to be diagonal {\it apriori}. Also,
note that we have already introduced the correct form of the spinor mass
term according to the prescription in eq.~(\ref{mass-dirac}) so that
this theory contains a Dirac spinor of mass $\mu$. The
theory~(\ref{adsmass}) can be shown to be invariant under the following
supersymmetry transformation rules:
\begin{equation} \label{susymass}
\begin{array}{l}
\delta \phi^{I} = (\sigma^{I} \epsilon)_{ij} \overline{\epsilon}^{i}
\lambda_{j} \\
\delta \lambda_{i} = i (\epsilon \sigma^{I})_{ji} \dirsl{\partial}
\phi^{I} \epsilon_{j} + a \frac{3}{2} ({\sigma^{I}}^{T} \epsilon
M)_{ij} \epsilon_{j} \phi^{I} + a \mu (M \epsilon
{\sigma^{I}}^{\dagger})_{ij} \epsilon_{j} \phi^{I},
\end{array}
\end{equation}
provided that the scalar mass matrix, $m^{2}_{IJ}$, takes a very
specific form, which will be determined by the supersymmetry. Obtaining
$m^{2}_{IJ}$ is nontrivial, so we provide the necessary calculations below.

Using the supersymmetry transformation rules~(\ref{susymass}) to vary
the action, we find that all the terms proportional to 1 and $a$ cancel
but terms proportional to $a^{2}$ yield the following matrix equation
for each $I$:
\begin{equation} \label{scalarmasseq}
\left(\mu^{2} - \frac{15}{4}\right) \sigma^{I} \epsilon + \mu M
\sigma^{I} \epsilon M = m^{2}_{IJ} \sigma^{J} \epsilon.
\end{equation}
One way to solve this equation is to expand everything in $\{
\sigma^{I}\}$ basis and then set the coefficients of each $\sigma^{I}$
matrix to 0. Noting that $M \sigma^{4} M = \sigma^{4}$, for $I = 1, 2,
3$ we write
\begin{equation} \label{Msigma}
M \sigma^{I} M = c_{IJ} \sigma^{J}
\end{equation}
where $c$ is a $3 \times 3$ matrix easily found from eq.~(\ref{M}):
\begin{equation}
\left( \begin{array}{ccc} 2 x_{1}^{2}-1 & 2 x_{1} x_{2} & 2 x_{1}
x_{3} \\ 2 x_{1} x_{2} & 2 x_{2}^{2}-1 & 2 x_{2} x_{3} \\ 2 x_{1}
x_{3} & 2 x_{2} x_{3} & 2 x_{3}^{2}-1 \end{array} \right)
\end{equation}
with \[ \vec{x} = (\sin\theta \cos\phi, \sin\theta \sin\phi,
\cos\theta). \] We note that for $I=2$ eq.~(\ref{scalarmasseq})
decouples from the rest of the equations giving us
\begin{equation}
m^{2}_{12} = m^{2}_{23} = m^{2}_{24} = 0, m^{2}_{22} = \mu^{2} + \mu -
\frac{15}{4}. 
\end{equation}
The other 3 equations remain coupled:
\begin{equation} \label{scalarmass1}
\begin{array}{l}
-(\mu^{2} - \frac{15}{4} - m^{2}_{11}) \sigma^{3} = \mu c_{3J}
\sigma^{J} + m^{2}_{13} \sigma^{1} - m^{2}_{14} \sigma^{2} \\
-(\mu^{2} - \frac{15}{4} - m^{2}_{33}) \sigma^{1} = \mu c_{1J}
\sigma^{J} + m^{2}_{13} \sigma^{3} + m^{2}_{34} \sigma^{2} \\
-(\mu^{2} - \frac{15}{4} - m^{2}_{44}) \sigma^{2} = \mu c_{2J}
\sigma^{J} + m^{2}_{34} \sigma^{1} - m^{2}_{14} \sigma^{3}
\end{array}
\end{equation}
where we used eq.~(\ref{Msigma}) above. A few simple manipulations yield
the values of the diagonal elements of the scalar mass matrix
\begin{equation} \label{scalarmass2}
\begin{array}{l}
m^{2}_{11} = \mu^{2} - \frac{15}{4} + \mu c_{33} \\
m^{2}_{33} = \mu^{2} - \frac{15}{4} + \mu c_{11} \\
m^{2}_{44} = \mu^{2} - \frac{15}{4} + \mu c_{22}.
\end{array}
\end{equation}
Plugging these values back into eq.~(\ref{scalarmass1}) we find all the
other elements of the scalar mass matrix:
\begin{equation} \label{scalarmass3}
m^{2}_{13} = - \mu c_{13}, ~ m^{2}_{14} = \mu c_{23}, ~ m^{2}_{34} = -
\mu c_{12}. 
\end{equation}
Finally, putting eqs.~(\ref{scalarmass2}) and (\ref{scalarmass3})
together we find the scalar mass matrix
\begin{equation} \label{m2}
m^{2} = \left(
\begin{array}{cccc}
\mu^{2} - \frac{15}{4} + \mu c_{33} & 0 & - \mu c_{13} & \mu c_{23} \\
0 & \mu^{2} + \mu - \frac{15}{4} & 0 & 0 \\
- \mu c_{13} & 0 & \mu^{2} - \frac{15}{4} + \mu c_{11} & - \mu c_{12} \\
\mu c_{23} & 0 & - \mu c_{12} & \mu^{2} - \frac{15}{4} + \mu c_{22} 
\end{array}
\right)
\end{equation}
where the rows and columns can be rearranged to give a block diagonal
form. To find the physical values of the masses we need to diagonalize
$m^{2}$ and read the physical masses off of the diagonal. This procedure
yields
\begin{equation} \label{m2phys}
m^{2} = \mu^{2} + \mu - \frac{15}{4}, ~ \mu^{2} + \mu - \frac{15}{4}, ~
\mu^{2} - \mu - \frac{15}{4}, ~ \mu^{2} - \mu - \frac{15}{4}.
\end{equation}
This answer is remarkable because these are precisely the masses we
expect from the AdS/CFT correspondence \cite{witten,sweedes}: from the
AdS/CFT correspondence we know that this theory should contain a complex
scalar with conformal dimension $\Delta$, a complex spinor with
conformal dimension $\Delta + \frac{1}{2}$, and another complex scalar
with conformal dimension $\Delta + 1$, which in 5 dimensions gives the
spinor mass 
\begin{equation}
\mu = \Delta + \frac{1}{2} - 2 = \Delta - \frac{3}{2}
\end{equation}
and the two scalar masses
\begin{equation} \begin{array}{l}
m^{2} = \Delta (\Delta - 4) \\ m^{2} = (\Delta+1) (\Delta - 3)
\end{array} \end{equation}
which in turn implies that the complex scalars in this theory should
have their masses equal to
\begin{equation} \label{m2corr}
\begin{array}{l}
m^{2} = \left( \mu + \frac{3}{2} \right) \left( \mu - \frac{5}{2}
\right) = \mu^{2} - \mu - \frac{15}{4} \\
m^{2} = \left( \mu + \frac{5}{2} \right) \left( \mu - \frac{3}{2}
\right) = \mu^{2} + \mu - \frac{15}{4}.
\end{array}
\end{equation}
Hence, eq.~(\ref{m2}) is the correct scalar mass matrix as the
equations~(\ref{m2phys}) and (\ref{m2corr}) are in exact agreement.

To complete the study of this theory we calculate the supersymmetry
algebra for the spinors and scalars under the transformation
rules~(\ref{susymass}). For the scalars, we obtain 
\begin{equation} \label{susyalgscmass}
\begin{array}{ll}
[\delta_{1},\delta_{2}] \phi^{I} & = 2 i \overline{\epsilon}_{1}^{i}
\gamma^{\mu} \epsilon_{2 i} \partial_{\mu} \phi^{I} \\ & + a \frac{3}{2}
\overline{\epsilon}_{2}^{i} \left(\sigma^{I} {\sigma^{J}}^{\dagger} M - 
M \sigma^{J} {\sigma^{I}}^{\dagger}\right)_{ij} \epsilon_{1 j} \phi^{J} \\ & +
a \mu \overline{\epsilon}_{2}^{i} \left(\sigma^{I} M^{\star}
{\sigma^{J}}^{\dagger} - \sigma^{J} M^{\star}
{\sigma^{I}}^{\dagger}\right)_{ij} \epsilon_{1 j} \phi^{J}
\end{array}
\end{equation}
which is almost the same as the supersymmetry algebra for the
``conformal scalar'' theory, eq.~(\ref{susyalgsc}), with addition of a
term proportional to the spinor mass, $\mu$. Because the scalar mass
term (\ref{m2}) breaks SO(4) symmetry down to SO(2)$\times$SO(2) (this fact
becomes obvious once the scalar fields are rotated so that the scalar
mass term becomes diagonal -- see discussion below), the two
non-derivative terms in the algebra must encode at least some part of this
symmetry of the Lagrangian. To see this more clearly, let us assume that
$\phi^{1}$, $\phi^{2}$ and $\phi^{3}$, $\phi^{4}$ have physical masses
$\mu^{2} + \mu - \frac{15}{4}$ and $\mu^{2} - \mu - \frac{15}{4}$
respectively, so that each pair transforms under separate
symmetries. First, note that even with this symmetry breaking, the
transformation
\begin{equation}
\delta \phi^{I} = \overline{\epsilon}_{2}^{i} \left(\sigma^{I} M^{\star}
{\sigma^{J}}^{\dagger} - \sigma^{J} M^{\star}
{\sigma^{I}}^{\dagger}\right)_{ij} \epsilon_{1 j} \phi^{J}
\end{equation}
is by itself a symmetry of the kinetic part of the scalar Lagrangian
because before the transformation (\ref{extraO4}) was a symmetry of the
kinetic part as well. To establish this fact is nontrivial, but it all
boils down to showing that
\begin{equation}
\left[ \sigma^{I} M^{\star} {\sigma^{J}}^{\dagger}, M \right] - \left[
\sigma^{J} M^{\star} {\sigma^{I}}^{\dagger}, M \right] = 0
\end{equation}
for all values of $I, J$. We can again rewrite the above transformation in a
more compact form, $\delta \phi^{I} = i \overline{\epsilon}_{2}^{i}
\epsilon_{1 i} Q^{IJ} \phi^{J}$ where $Q$ is a $4 \times 4$ matrix
\begin{equation} \label{Q}
Q = \left( \begin{array}{cccc} 0 & -x_{3} & -x_{2} & -x_{1} \\ x_{3} & 0 &
-x_{1} & x_{2} \\ x_{2} & x_{1} & 0 & -x_{3} \\ x_{1} & -x_{2} & x_{3} & 0
\end{array} \right).
\end{equation}
Now, we only need to understand how the transformation
(\ref{susyalgscmass}) acts on the mass term. To see this more clearly,
let us redefine
\begin{equation}
{\phi^{\prime}}^{I} = O^{IJ} \phi^{J}
\end{equation}
with the matrix $O$ defined as follows:
\begin{equation} \label{O}
O = \left( \begin{array}{cccc} x_{3} & 0 & \sqrt{1-x_{3}^{2}} & 0 \\ 0 &
1 & 0 & 0 \\ -x_{1} & 0 & \frac{x_{1} x_{3}}{\sqrt{1-x_{3}^{2}}} &
\frac{x_{2}}{\sqrt{1-x_{3}^{2}}} \\ x_{2} & 0 & -\frac{x_{2}
x_{3}}{\sqrt{1-x_{3}^{2}}} & \frac{x_{1}}{\sqrt{1-x_{3}^{2}}}
\end{array} \right).
\end{equation}
With this field redefinition, the scalar mass matrix becomes diagonal in
precisely the way discussed above and the transformation
(\ref{susyalgscmass}) becomes
\begin{equation} \label{O4-2O2}
\delta {\phi^{\prime}}^{I} = - i \overline{\epsilon}_{2}^{i}
\gamma^{\mu} \epsilon_{1 i} \partial_{\mu} {\phi^{\prime}}^{I} + i a
\overline{\epsilon}_{2}^{i} \epsilon_{1 i} \left[ 3 (O^{T} T O)^{IJ} +
2 \mu (O^{T} Q O)^{IJ} \right] {\phi^{\prime}}^{J}
\end{equation}
where $T$ and $Q$ are defined in eqs.~(\ref{T}) and (\ref{Q})
respectively. Using eq.~(\ref{O}) we can explicitely compute
\begin{equation}
O^{T} T O = \left( \begin{array}{cccc} 0 & 1 & 0 & 0 \\ -1 & 0 &
0 & 0 \\ 0 & 0 & 0 & -1 \\ 0 & 0 & 1 & 0 \end{array} \right), ~ 
O^{T} Q O = \left( \begin{array}{cccc} 0 & -1 & 0 & 0 \\ 1 & 0 &
0 & 0 \\ 0 & 0 & 0 & -1 \\ 0 & 0 & 1 & 0 \end{array} \right).
\end{equation}
Hence the transformation (\ref{O4-2O2}) of the new fields encodes a
particular SO(2) symmetry of the larger SO(2)$\times$SO(2) symmetry of the
scalar Lagrangian.

The spinor supersymmetry algebra is quite similar to that of the
previously derived conformal scalar theory, eq.~(\ref{susyalgsp}):
\begin{equation} \label{susyalgspmass}
[\delta_{1},\delta_{2}] \lambda_{i} = 2 i D_{\mu} \lambda_{i}
\overline{\epsilon}_{1}^{j} \gamma^{\mu} \epsilon_{2 j} - 2 a \mu M_{ik}
\lambda_{k} \overline{\epsilon}_{1}^{j} \epsilon_{2 j} + \frac{a}{2}
\gamma^{\mu \nu} \lambda_{i} M_{kj} \overline{\epsilon}_{1}^{k}
\gamma_{\mu \nu} \epsilon_{2 j}
\end{equation}
with a new term proportional to the spinor mass $\mu$. As before, the
third term with $\gamma_{\mu\nu}$ in this algebra is exactly the extra
symmetry term required to recover the O(4,2) symmetry of the Lagrangian,
and the new term proportional to $\mu$ is reminiscent of the U(1)
symmetry (\ref{extraspinsym}) in the Yang-Mills theory described in
Section 3.

\sect{Acknowledgments}

I would like to thank Prof. D.Z. Freedman for suggesting this problem to
me, as well as for his help, advice, guidance, and helpful
discussions. I would also like to thank A. Zaffaroni for his fruitful
discussions with me. Research is supported in part by funds provided by
the National Science Foundation (NSF) through the NSF Graduate
Fellowship and by the U.S. Department of Energy (D.O.E.) under
cooperative research agreement \#DE-FC02-94ER40818.

\appendix
\begin{center} {\Large {\bf Appendix}} \end{center}
\sect{Summary of results in $(-,+, \cdots ,+)$ signature}

In this section we will summarize some of the formulas given in body of
the text for the $(-,+, \cdots ,+)$ signature. We do so because most of the
recent literature on AdS uses this signature almost exclusively. Note
that in this signature, the curvature takes the usual form, $R_{\mu\nu}
= - (d - 1) a^{2} g_{\mu\nu}$.

Most of the formulas can be converted to the $(-,+, \cdots ,+)$
signature simply by changing $\gamma_{\mu} \rightarrow -i \gamma_{\mu}$
for every $\gamma_{\mu}$ in the formula. Hence, a complex unconstrained
Killing spinor 
equation~(\ref{kilspdirac}) becomes:
\begin{equation}
D_{\mu} \epsilon = \frac{a}{2} \gamma_{\mu} \epsilon,
\end{equation}
its solution (\ref{kilspn}) becomes:
\begin{equation}
\epsilon = e^{\frac{1}{2} a r \gamma_{r}} \left( 1 + \frac{1}{2}
a x^{\alpha} \gamma_{\alpha} (1 - \gamma_{r}) \right) \epsilon_{0},
\end{equation}
symplectic Majorana Killing spinor equation~(\ref{kilspeq}) becomes:
\begin{equation}
D_{\mu} \epsilon_{i} = M_{i j} \frac{a}{2} \gamma_{\mu} \epsilon_{j},
\end{equation}
and its solution (\ref{kilspM}) becomes:
\begin{equation}
\epsilon_{i} = \left(e^{\frac{1}{2} a r M \gamma_{r}}\right)_{ij} \left(
\delta_{jk} + \frac{1}{2} a x^{\alpha} \gamma_{\alpha} (M_{jk} - 
\delta_{jk} \gamma_{r}) \right) \xi_{k}.
\end{equation}
Other formulas, such as supersymmetry transformation rules, have to be
checked carefully when changing signatures so that the properties of the
fields, e.g. real or symplectic Majorana, are satisfied by the
transformations. When this is done carefully, we find that the
Yang-Mills theory transformation rules (\ref{susygauge}) become:
\begin{equation}
\begin{array}{l}
\delta A_{\mu} = - i \overline{\eta}^{i} \gamma_{\mu} \chi_{i} \\
\delta \phi = \overline{\eta}^{i} \chi_{i} \\
\delta \chi_{i} = (-\sigma_{\mu\nu} F^{\mu\nu} + i \dsl{D} \phi)
\eta_{i} + 2 i a \phi M_{ij} \eta_{j},
\end{array}
\end{equation}
the conformal scalar theory transformation rules (\ref{susyconf}) become:
\begin{equation}
\begin{array}{l}
\delta \phi^{I} = - i (\sigma^{I} \epsilon)_{ij} \overline{\epsilon}^{i}
\lambda_{j} \\
\delta \lambda_{i} = (\epsilon \sigma^{I})_{ji} \dirsl{\partial}
\phi^{I} \epsilon_{j} - a \frac{3}{2} ({\sigma^{I}}^{T} \epsilon M)_{ij}
\epsilon_{j} \phi^{I},
\end{array}
\end{equation}
and the massive scalar theory transformation rules (\ref{susymass}) become:
\begin{equation}
\begin{array}{l}
\delta \phi^{I} = -i (\sigma^{I} \epsilon)_{ij} \overline{\epsilon}^{i}
\lambda_{j} \\
\delta \lambda_{i} = (\epsilon \sigma^{I})_{ji} \dirsl{\partial}
\phi^{I} \epsilon_{j} - a \frac{3}{2} ({\sigma^{I}}^{T} \epsilon
M)_{ij} \epsilon_{j} \phi^{I} + i a \mu (M \epsilon
{\sigma^{I}}^{\dagger})_{ij} \epsilon_{j} \phi^{I}.
\end{array}
\end{equation}

\sect{Isometry transformations for spinors on AdS}

In this section we will attempt to prove that the action of a free,
massless spinor on AdS$_{d}$ is invariant under
\begin{equation}
\delta \psi = K^{\mu} D_{\mu} \psi + \frac{1}{4} D^{\mu} K^{\nu}
\gamma_{\mu \nu} \psi
\end{equation}
where $K^{\mu}$ is an O($d-1,2$) Killing vector. This proof remains true
in any spacetime dimension and for any spinor whose action is given by
the usual Lagrangian
\begin{equation}
{\cal L} = i \overline{\psi} \dsl{D} \psi.
\end{equation}

Although O($d-1,2$) Killing vectors and their properties on AdS$_{d}$
can be studied independently of the Killing spinors, we will use the
definition of Killing vectors through Killing spinors (true only in some
dimensions) as a shortcut to establish the following property of the
Killing vectors:
\begin{equation} \label{DDK}
D_{\mu} D_{\nu} K_{\rho} = D_{\mu} D_{\nu} \left( i
\overline{\epsilon}_{1} \gamma_{\rho} \epsilon_{2}\right) = a D_{\mu}
\left(\overline{\epsilon}_{1} \gamma_{\nu\rho} \epsilon_{2}\right) =
a^{2} \left(g_{\mu\rho} K_{\nu} - g_{\mu\nu} K_{\rho}\right)
\end{equation}
where in the intermediate steps we used the Killing spinor
equation~(\ref{kilspdirac}) and the properties of the Clifford algebra
\begin{equation}
\{\gamma_{\mu},\gamma_{\nu}\} = 2 g_{\mu\nu}.
\end{equation}
Note that although the intermediate steps in eq.~(\ref{DDK}) involve
Killing spinors, the final result is expressed only in terms of the
Killing vectors. Hence, this is a general property of the O($d-1,2$)
Killing vectors. Similarly, we can establish another Killing vector
property:
\begin{equation}
D_{\mu} K_{\nu} = - D_{\nu} K_{\mu}.
\end{equation}

Now, let us vary the free action by
\begin{equation}
\delta_{1} \psi = K^{\mu} D_{\mu} \psi
\end{equation}
which yields:
\begin{equation} \begin{array}{ll}
\delta_{1} \left(\overline{\psi} \dsl{D} \psi \right) & = -
\overline{\psi} K^{\mu} D_{\mu} \dsl{D} \psi + \overline{\psi} K^{\mu}
\dsl{D} D_{\mu} \psi + \overline{\psi} \gamma_{\nu} D_{\mu} \psi D^{\nu}
K^{\mu} \\ & = \overline{\psi} \gamma^{\nu} [D_{\nu}, D_{\mu}] \psi
K^{\mu} + \overline{\psi} \gamma_{\nu} D_{\mu} \psi D^{\nu} K^{\mu} \\ &
= a^{2} \frac{d-1}{2} \overline{\psi} \gamma_{\mu} \psi K^{\mu} +
\overline{\psi} \gamma_{\nu} D_{\mu} \psi D^{\nu} K^{\mu},
\end{array} \end{equation}
where to go from line 2 to line 3 we used the Ricci identity,
eq.~(\ref{ricci}). Thus, it is clear that the above transformation is
not a symmetry of the Lagrangian. Now, let us vary the action by
\begin{equation}
\delta_{2} \psi = D^{\mu} K^{\nu} \gamma_{\mu\nu}\psi
\end{equation}
which gives
\begin{equation} \begin{array}{ll}
\delta_{2} \left(\overline{\psi} \dsl{D} \psi \right) & = - D_{\mu}
K_{\nu} \overline{\psi} \gamma^{\mu\nu} \dsl{D} \psi + D_{\mu} K_{\nu}
\overline{\psi} \dsl{D} \gamma^{\mu\nu} \psi + D_{\rho} D_{\mu} K_{\nu}
\overline{\psi} \gamma^{\rho} \gamma^{\mu\nu}\psi \\ & = 2 D_{\mu} K_{\nu}
\overline{\psi} (\gamma^{\nu} g^{\mu\rho} - \gamma^{\mu} g^{\nu\rho})
D_{\rho} \psi - 2 a^{2} (d - 1) \overline{\psi} \gamma_{\mu} \psi K^{\mu}
\\ & = - 4 \overline{\psi} \gamma_{\nu} D_{\mu} \psi D^{\nu} K^{\mu} - 2
a^{2} (d - 1) \overline{\psi} \gamma_{\mu} \psi K^{\mu}
\end{array} \end{equation}
where to go from line 1 to line 2 we used the properties of the Clifford
algebra and of the Killing vectors, eq.~(\ref{DDK}). Therefore, it is
now clear that the action of a free, massless spinor on AdS is invariant
under
\begin{equation}
(\delta_{1} + \delta_{2}) \psi = K^{\mu} D_{\mu} \psi + \frac{1}{4}
D^{\mu} K^{\nu} \gamma_{\mu \nu} \psi.
\end{equation}
Let us finally note that for $d = 5$, the preceding proof can be
applied verbatim to the case of the symplectic Majorana spinors if we
note that on AdS$_{5}$ 
\begin{equation} \begin{array}{l}
K^{\mu} = i \overline{\epsilon}_{1}^{i} \gamma^{\mu} \epsilon_{2 i} \\
D_{\mu} K_{\nu} = a M_{ij} \overline{\epsilon}_{1}^{i} \gamma_{\mu\nu}
\epsilon_{2 j} = - D_{\nu} K_{\mu} \\
D_{\mu} D_{\nu} K_{\rho} = a^{2} (g_{\mu\rho} K_{\nu} - g_{\mu\nu} K_{\rho}).
\end{array}
\end{equation}

\sect{Results in $d \neq 5$}

In this section, we give some results for dimensions other than 5. These
results hold for those dimensions where symplectic Majorana spinors can
be defined and the charge conjugation matrix can be chosen to satisfy
eq.~(\ref{chargeconj}). We know \cite{pvan} that both conditions can be
satisfied in $d = 5, 6$ mod 8. Also, it is conceivable that similar
approach has to be taken for $d = 8, 9$ mod 8 as the only way to define
Majorana spinors there is to take a symmetric charge conjugation matrix
that satisfies eq.~(\ref{chargeconj}). However, it is important to
realize that the transformation rules given below do not describe
supersymmetry in dimensions above 5, but instead describe some
accidental symmetry of the free non-interacting Lagrangian. We give the
transformation rules and particle masses consistent with these
transformations for the theories discussed in the paper. Note that the
spinor algebra following from these transformations will
change because Fierz identities take different from in different dimensions.

{\bf Yang-Mills theory} is invariant under
\begin{equation}
\begin{array}{l}
\delta A_{\mu} = i \overline{\eta}^{i} \gamma_{\mu} \chi_{i} \\
\delta \phi = i \overline{\eta}^{i} \chi_{i} \\
\delta \chi_{i} = (\sigma_{\mu\nu} F^{\mu\nu} - \dsl{D} \phi) \eta_{i} -
i a (d - 3) \phi M_{ij} \eta_{j} 
\end{array}
\end{equation}
with mass parameters
\begin{equation} 
\mu = - \frac{d - 4}{4} M, ~ m^{2} = - 2 (d -3).
\end{equation}

{\bf Conformal scalar theory} is invariant under
\begin{equation}
\begin{array}{l}
\delta \phi^{I} = (\sigma^{I} \epsilon)_{ij} \overline{\epsilon}^{i}
\lambda_{j} \\
\delta \lambda_{i} = i (\epsilon \sigma^{I})_{ji} \dirsl{\partial}
\phi^{I} \epsilon_{j} + a \frac{d - 2}{2} ({\sigma^{I}}^{T} \epsilon M)_{ij}
\epsilon_{j} \phi^{I}
\end{array}
\end{equation}
with mass of the scalars given by
\begin{equation}
m^{2} = - \frac{d(d - 2)}{4},
\end{equation}
which is exactly the mass of the conformally coupled scalar in dimension
$d$ \cite{burges,mezincescu}.

{\bf Massive scalar theory} is invariant under
\begin{equation}
\begin{array}{l}
\delta \phi^{I} = (\sigma^{I} \epsilon)_{ij} \overline{\epsilon}^{i}
\lambda_{j} \\
\delta \lambda_{i} = i (\epsilon \sigma^{I})_{ji} \dirsl{\partial}
\phi^{I} \epsilon_{j} + a \frac{d - 2}{2} ({\sigma^{I}}^{T} \epsilon
M)_{ij} \epsilon_{j} \phi^{I} + a \mu (M \epsilon
{\sigma^{I}}^{\dagger})_{ij} \epsilon_{j} \phi^{I}
\end{array}
\end{equation}
with the scalar masses given by
\begin{equation}
m^{2} = \mu^{2} + \mu - \frac{d(d-2)}{4}, ~ \mu^{2} - \mu - \frac{d(d-2)}{4}
\end{equation}
each with multiplicity 2.


\begin{thebibliography}{99}
 \bibitem{maldacena} Juan Maldacena, {\it The large {$N$} limit of
 superconformal field theories and supergravity},
 Adv. Theor. Math. Phys. {\bf 2} (1998) 231, hep-th/9711200.
 \bibitem{klebanov} S.S. Gubser, I.R. Klebanov and A.M. Polyakov, {\it
 Gauge theory correlators from noncritical string theory},
 Phys. Lett. {\bf B428}, (1998) 105, hep-th/9802109.
 \bibitem{witten} Edward Witten, {\it Anti-de Sitter space and
 holography}, Adv. Theor. Math. Phys. {\bf 2} (1998) 253-291,
 hep-th/9802150.
 \bibitem{pvan} P. Van Nieuwenhuizen, {\it An introduction to simple
 supergravity and the Kaluza-Klein program} in {\bf Les Houches 1983,
 Proceedings, Relativity, Groups and Topology, Ii}, 823-932.
 \bibitem{cremmer} E. Cremmer, {\it Supergravities in 5 dimensions} in
 {\bf Superspace and Supergravity, Proceedings of the Nuffield Workshop}
 edited by S.W. Hawking and M. Ro{\v{c}}ek, 267-282.
 \bibitem{gunaydin} M. G{\"{u}}naydin, L.J. Romans, and N.P. Warner,
 {\it Compact and non-compact gauged supergravity theories in five
 dimensions}, Nucl. Phys. {\bf B272} (1986) 598.
 \bibitem{burges} Christopher J.C. Burges, Daniel Z. Freedman, S. Davis,
 G.W. Gibbons, {\it Supersymmetry in Anti-de Sitter space}, Annals
 Phys. {\bf 167} (1986) 285.
 \bibitem{lupope} H. L{\"{u}}, C.N. Pope, and P.K. Towsend, {\it Domain walls
 from Anti-de Sitter spacetime}, Phys. Lett. {\bf B391} (1997) 39,
 hep-th/9607164.
 \bibitem{lupoper} H. L{\"{u}}, C.N. Pope, and J. Rahmfeld, {\it A
 construction of killing spinors on S{$^{N}$}}, hep-th/9805151.
 \bibitem{zizzi} P.A. Zizzi, {\it The USp(2) Supersymmetric SU(2)
 Yang-Mills theory in 4+1 dimensions and the central charge},
 Nucl. Phys. {\bf B189} (1981) 317.
 \bibitem{kimromans} H.J. Kim, L.J. Romans, and P. van Nieuwenhuizen,
 {\it Mass spectrum of chiral ten-dimensional N=2 supergravity on
 S{$^{5}$}}, Phys. Rev. {\bf D32} (1985) 389.
 \bibitem{sweedes} Mans Henningson, Konstadinos Sfetsos, {\it Spinors
 and the AdS/CFT correspondence}, Phys. Lett. {\bf B431} (1998) 63-68,
 hep-th/9803251.
 \bibitem{sharpe} Eric Sharpe, {\it Boundary superpotentials},
 Nucl. Phys. {\bf B523} (1998) 211, hep-th/9611196.
 \bibitem{mezincescu} L. Mezincescu and P.K. Townsend, {\it Stability At
 A Local Maximum In Higher Dimensional Anti-De Sitter Space And
 Applications To Supergravity}, Ann. Phys. {\bf 160}, (1985) 406.
\end{thebibliography}
\end{document}